\documentclass{tCPH2e-modified}
\usepackage[utf8]{inputenc}
\begin{document}


\jvol{} \jnum{} \jyear{} \jmonth{}

\markboth{S. Schippers, A. L. D. Kilcoyne, R. A. Phaneuf, A. M\"{u}ller}{Photoionization of ions with synchrotron radiation: From ions in space to atoms in cages}


\title{Photoionization of ions with synchrotron radiation:\\ From ions in space to atoms in cages}

\author{Stefan Schippers$^{a}$$^{\ast}$\thanks{$^\ast$Corresponding author. Email: Stefan.Schippers@physik.uni-giessen.de
\vspace{6pt}}, A. L. David Kilcoyne$^b$, Ronald A. Phaneuf$^{\,c}$ and Alfred M\"{u}ller$^{d}$\\\vspace{6pt}
$^{a}${\em{I. Physikalisches Institut, Justus-Liebig-Universit\"{a}t Gie{\ss}en, 35392 Gie{\ss}en, Germany}}\\
$^{b}${\em{Advanced Light Source, Lawrence Berkeley National Laboratory, Berkeley, CA 94720, USA}}\\
$^{c}${\em{Department of Physics, University of Nevada, Reno, NV 89557-0220, USA}}\\
$^{d}${\em{Institut f\"{u}r Atom- und Molek\"{u}lphysik, Justus-Liebig-Universit\"{a}t Gie{\ss}en, 35392 Gie{\ss}en, Germany}}
}

\maketitle

\begin{abstract}
The photon-ion merged-beams technique for the photoionization of mass/charge selected ionized atoms, molecules and clusters by x-rays from synchrotron radiation sources is introduced. Examples for photoionization of atomic ions are discussed by going from outer-shell ionization of simple few-electron systems to inner-shell ionization of complex many-electron ions. Fundamental ionization mechanisms are elucidated and the importance of the results for applications in astrophysics and plasma physics is pointed out. Finally, the unique capabilities of the photon-ion merged-beams technique for the study of photoabsorption by nanoparticles are demonstrated by the example of endohedral fullerene ions.
\end{abstract}

\bigskip
\bigskip

\section{Introduction}

Photoionization of ions occurs in many cosmic environments where matter is in the plasma state. Examples are active galactic nuclei, hydrogen clouds around hot stars, and the interior of the Sun. Among other atomic quantities, accurate photoionization cross sections of atomic ions are required for being able to infer the physical conditions of these objects from astronomical observations \cite{Kallman2010}. An entire field of research, \lq\lq Laboratory Astrophysics\rq\rq\  \cite{Savin2012}, is devoted to providing the atomic data that are required for the understanding of cosmic nonequilibrium plasmas. The \lq\lq Opacity Project\rq\rq, for example, aims at calculating accurate cross sections for the photoionization of ions which are needed for estimating the opacity of stellar matter (see \cite{Badnell2005a} and references therein). In general, the atomic data needs for the modeling of nonequilibrium plasmas are vast. Most of the body of compiled data comes from theory which has to use approximations to make calculations tractable even with modern computer technology at hand. Therefore, benchmarking by experiment is vitally needed.

\begin{figure}[t]
\centering{\includegraphics[width=0.35\textwidth]{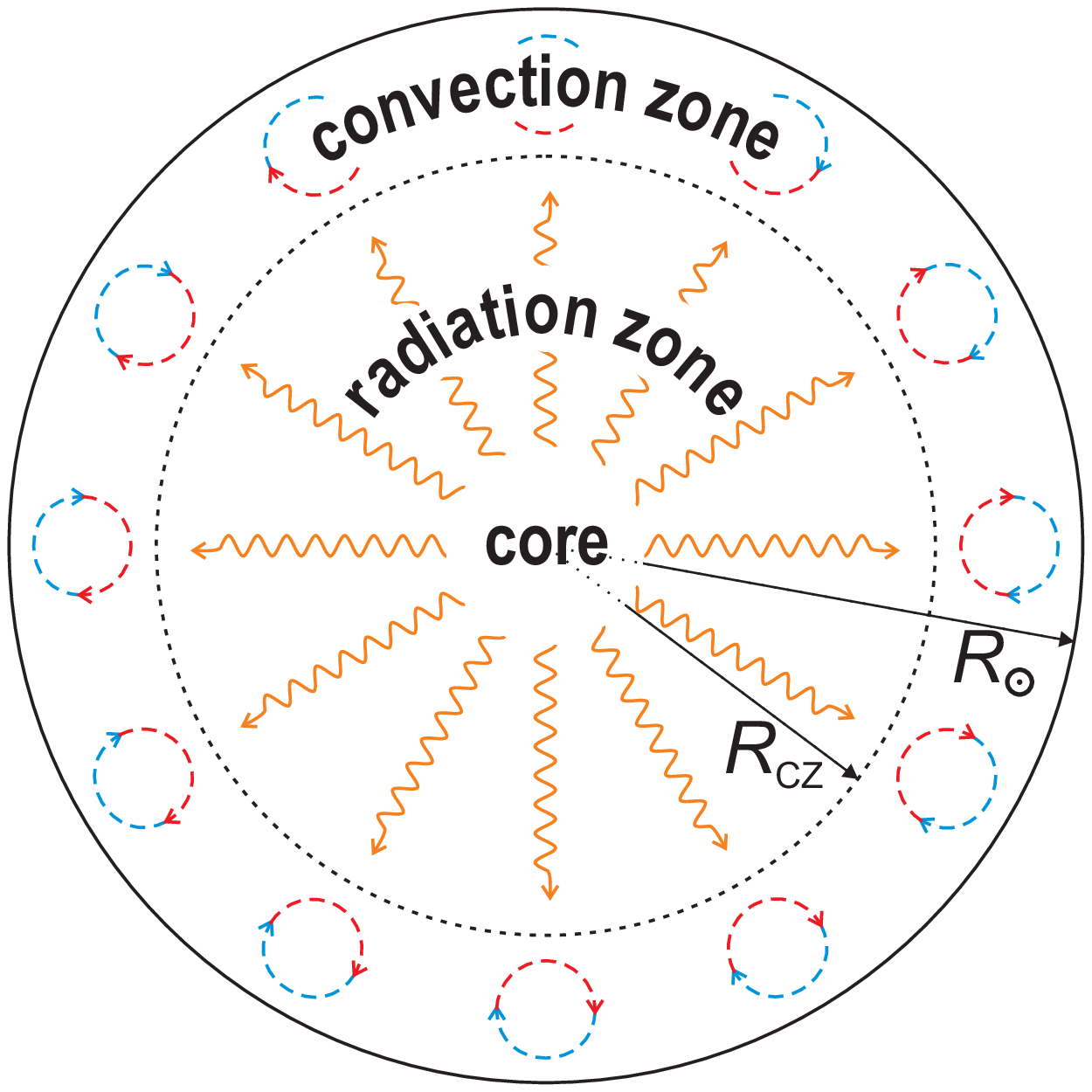}}
\caption{\label{fig:TheSun} Simplified shell structure of the Sun (with radius $R_{\odot}$). The dashed line marks the boundary (at radius $R_\mathrm{CZ}$) between the radiative zone and the convective zone. }
\end{figure}

A problem from solar physics illustrates this situation. The Sun produces energy by nuclear fusion reactions which proceed most efficiently at the high temperatures and pressures that prevail in the core of the Sun. The mechanism of energy transport to the Sun's surface depends on the radial distance from the core. At smaller radii, where the solar matter is largely ionized, energy is transported by radiation. At larger distances from the core, where the solar matter is less dense, energy is transported by convective mass flow of mainly neutral hydrogen. Figure \ref{fig:TheSun} visualizes the shell structure of the Sun. The radius $R_\mathrm{CZ}$ of the boundary between the radiation and the convection layers can be inferred from helioseismology \cite{Chaplin2005} and from solar models \cite{Bahcall1988} that quantitatively treat energy generation and transport in the Sun. Ingredients of the solar models are, among others, the photoionization cross sections from the Opacity Project.

\begin{figure}[t]
\centering{\includegraphics[width=0.5\textwidth]{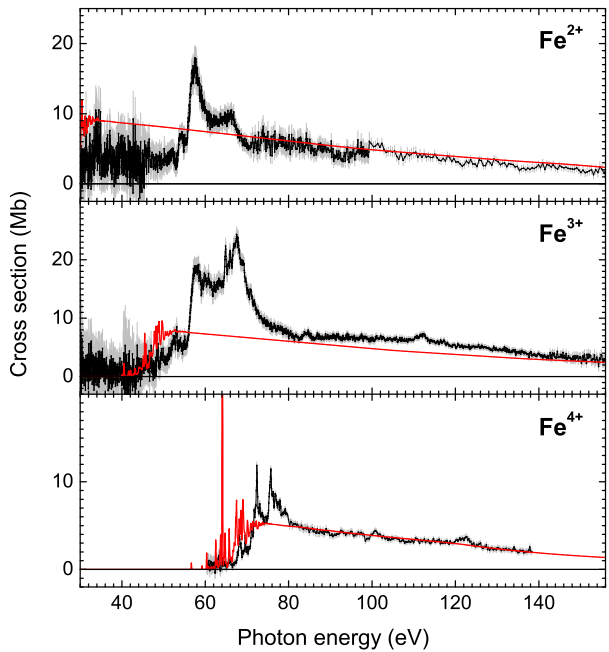}}
\caption{\label{fig:FeAstrid} (adapted with permission from Ref.~\cite{ElHassan2009}. Copyrighted by the
American Physical Society) Comparison of experimentally measured photoionization cross sections (black symbols with grey error bars) with theoretical data from the TOPbase (red full lines). The measurements were carried out at a photon-ion merged-beams setup located at the ASTRID synchrotron in Aarhus, Denmark.}
\end{figure}

The problem is now that the value $R_\mathrm{CZ}=0.713\pm0.001 R_\odot$ from helioseismology and the values of up to $R_\mathrm{CZ}=0.72 R_\odot$ from the latest solar models significantly differ from one another \cite[][and references therein]{Basu2014}. Similar discrepancies exist also for other solar quantities of interest as, e.g., the flux of neutrinos. It has been speculated \cite{Serenelli2009,Bailey2015} that part of the discrepancies might be caused by inaccurate photoionization cross sections. Theoretically derived photoionization cross sections from the Opacity Project are partly accessible via the \lq\lq TOPbase\rq\rq\ data base \cite{Cunto1993}. Figure \ref{fig:FeAstrid} compares experimentally measured cross sections for the photoionization of low-charged iron ions with the corresponding quantities from the TOPbase. The very significant differences between the experimental and the theoretical cross section curves clearly illustrate the need for experimental benchmarking.

The experimental data in Figure~\ref{fig:FeAstrid} were obtained with the photon-ion merged-beams technique which is explained in some more detail in section~\ref{sec:exp} of the present paper. Section~\ref{sec:atomic} presents and discusses selected results for atomic ions of astrophysical interest. In recent years, the photon-ion merged-beams technique has also been used for photoionization studies with cluster ions such as (endohedral) fullerene ions \cite[see, e.g.,][]{Phaneuf2013}. One particulary illustrative result for Xe@C$_{60}^+$ is presented in section~\ref{sec:full}. The present paper is intended to serve as an introduction for the non-specialized reader into the topic of  photoionization of ions. It does not provide a comprehensive overview over this field of research. For this, the reader is referred to the review articles by West \cite{West2001a}, Kennedy et al.\ \cite{Kennedy2004a},  Kjeldsen \cite{Kjeldsen2006a}, Berrah et al.\ \cite{Berrah2011} and Müller \cite{Mueller2015}.

\section{The photon-ion merged-beams technique}\label{sec:exp}

\begin{figure}[b]
\centering{\includegraphics[width=0.5\textwidth]{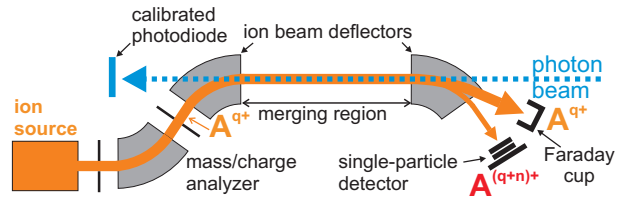}}
\caption{\label{fig:MB} Essential features of a photon-ion merged-beams arrangement. The reaction products A$^{(q+n)+}$  result from $n$-fold ionization of the A$^{q+}$ primary ions by photon impact.}
\end{figure}

Ionization energies \cite{Kramida2014} of positively charged atomic ions range from about 10 eV for Ba$^+$ up to 132 keV for hydrogen-like U$^{91+}$. Thus, experiments on photoionization of atomic ions require UV and X-ray photons. Sufficiently intense beams of such photons only became available with the advent of 2nd generation synchrotron light sources in the late 1970s. The first experiments on the photoionization of ions were carried out in the mid 1980s by Lyon et al.\ \cite{Lyon1986} who pioneered the photon-ion merged-beams technique at the Daresbury Synchroton Radiation Source in the UK. Since then, the technique has been implemented at other synchrotron radiation sources around the world including ASTRID (Denmark) \cite{Kjeldsen1999b}, Spring-8 (Japan) \cite{Yamaoka2001}, ALS (USA) \cite{Covington2002},  Soleil (France) \cite{Gharaibeh2011a}, and PETRA\,III (Germany) \cite{Schippers2014}.

A particular difficulty associated with ionic targets is their extreme diluteness. The number of ions per unit volume is limited by the mutual electrostatic repulsion of the ions. Typical ion number densities in an ion beam are about $10^6$~cm$^{-3}$. This is 10 orders of magnitude smaller as compared to the number density of neutral atoms or molecules in a gas target with a pressure of 1~mbar. The photon-ion merged-beams technique makes up for the diluteness of the ionic target by providing a large interaction volume and a high efficiency for the detection of product ions.

\begin{figure}
\centering{\includegraphics[width=0.8\textwidth]{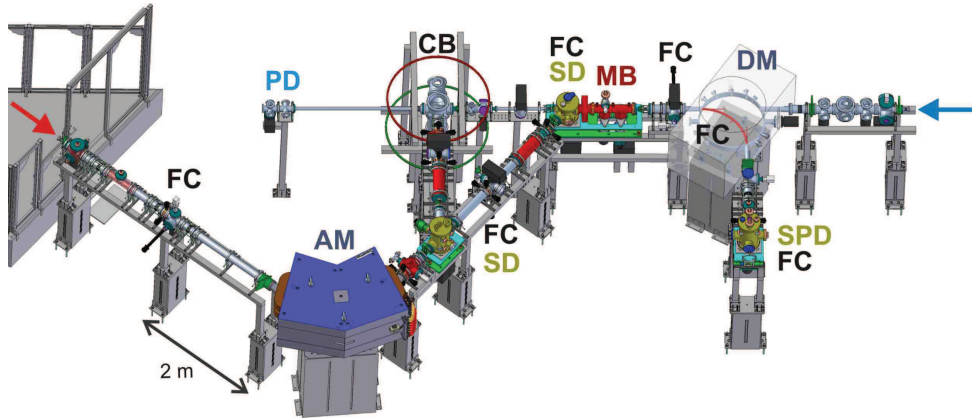}}
\caption{\label{fig:PIPE}(taken from Ref.~\cite{Schippers2014}. \copyright\ 2014 IOP Publishing.  Reproduced with permission.  All rights reserved) Sketch of the \underline{P}hoton-\underline{I}on spectrometer at \underline{PE}TRA\,III (PIPE), a permanent photon-ion end station at the \lq\lq Variable Polarization XUV Beam Line\rq\rq\ (P04) \cite{Viefhaus2013} of the PETRA\,III synchrotron radiation source operated by DESY in Hamburg, Germany. The photon beam enters the setup from the right (blue arrow). It is stopped by a calibrated photodiode (PD) which continuously monitors the absolute photon flux. The ion beam enters from the left (red arrow). It is generated with an ion source that is mounted on a separate platform (not fully shown). The analyzing magnet (AM) provides mass/charge selection of ions for further ion-beam transport. Spherical deflectors (SD) can be used to direct the ions either into the crossed-beams (CB) interaction point or into the merged-beams (MB) collinear beam overlap region. The demerging magnet (DM) deflects primary and product ions out of the photon-beam axis and directs product ions into the single-particle detector (SPD). The ion current can be measured at various places along the ion beamline by inserting Faraday cups (FC) into the ion beam. One Faraday cup is mounted inside the demerging magnet such that product ions which are deflected by $90^\circ$ can pass towards the single-particle detector, while ions in different charge states or with different kinetic energies are collected in this Faraday cup. The merged-beams interaction region is equipped with scanning slits for beam-profile measurements. }
\end{figure}

\begin{figure}[b]
\centering{\includegraphics[width=0.5\textwidth]{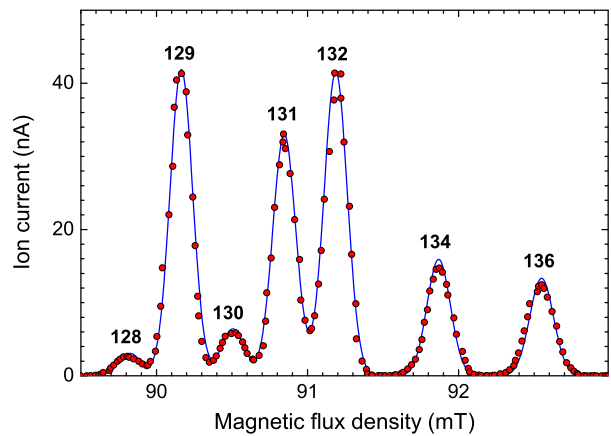}}
\caption{\label{fig:Xe2mass} (Modified version of a Figure from Ref.~\cite{Schippers2014}. \copyright\ 2014 IOP\ Publishing.  Reproduced with permission.  All rights reserved) Isotopic composition of a Xe$^{2+}$ ion beam as revealed by scanning the magnetic field of the beam analyzing magnet. The peaks are labelled by the atomic mass number of the corresponding xenon isotope. The symbols represent the result of the experimental mass scan. The full line has been calculated from the natural abundances of the xenon isotopes assuming a mass resolving power $m/\Delta m= 250$. }
\end{figure}

Figure~\ref{fig:MB} sketches the salient features of the experimental arrangement and Figure~\ref{fig:PIPE} presents details of the layout of the PIPE facility \cite{Schippers2014} at PETRA\,III. A beam of $q$-fold charged ions (denoted by A$^{q+}$) is produced by an ion source which is kept on an acceleration potential $U_\mathrm{acc}$ of a few kV. Thus, after acceleration, the ion kinetic energy is $E_\mathrm{ion} = \frac{1}{2}m_\mathrm{ion}v_\mathrm{ion}^2 = qeU_\mathrm{acc}$ with $e$ denoting the elementary charge. The ion source usually delivers ions in different charge states. In addition, impurity ions from the residual gas in the ion source may be present in the beam. Therefore, a dipole bending magnet in combination with beam-size defining slits (labelled \lq mass/charge analyzer\rq\ in Figure~\ref{fig:MB}) is used for selecting ions of the desired mass-to-charge-ratio to be transported further to the photon-ion interaction region. The resolving power of such an arrangement can be adjusted (at the expense of beam intensity) by closing the beam-size defining slits. Even different isotopes of a given ion species can be individually selected if required (Figure~\ref{fig:Xe2mass}).

The large interaction volume is created by deflecting the ion beam onto the photon-beam axis. The length of the merging region is typically 1--2~m depending on ion-optical requirements and on the floor space available. The high efficiency for the detection of product ions is due to the use of a beam of fast ions as a target for photoionization. The photo ions that are created in the photon-ion interaction region move in the same direction as the primary ion beam. A second magnetic or electrostatic deflector that is located downstream of the interaction region demerges ion and photon beams and at the same time separates the more highly charged product ions from the primary ions. The primary beam current is measured with a Faraday cup. The photo ions are counted one by one in a single particle detector. When carefully designed, such a detector has an efficiency of practically 100\% \cite{Rinn1982,Spruck2015}.

The well defined beam-beam collision geometry of the merged-beams technique allows an accurate determination of absolute photoionization cross sections $\sigma$ (see, e.g., \cite{Schippers2014}) by normalizing the measured detector count rate $R$ on the photon flux $\phi_\mathrm{ph}$, the number of ions in the interaction region, the beam overlap, and detection efficiency $\eta$, i.e.,
\begin{equation}\label{eq:sigma}
 \sigma = R \frac{q\,e\,v_\mathrm{ion}}{\eta\,I_\mathrm{ion}\,\phi_\mathrm{ph}\,\mathcal{F}_L}
\end{equation}
where $I_\mathrm{ion}$ denotes the electrical ion current. The beam overlap factor $\mathcal{F}_L$ for the length $L$ of the beam overlap region is determined by separate beam-profile measurements using scanning slits or rotating wires. The systematic uncertainty of the measured absolute cross section amounts to typically 10--15\% with the largest contribution to the error budget stemming from the beam overlap measurement \cite[see, e.g.,][]{Covington2002}.

In principle, the photon-ion merged-beams technique enables studies of the photoionization of any ion including negatively charged species \cite{Berrah2011} and complex molecular ions (see Section \ref{sec:full}). The maximum charge state that can be investigated at a given setup depends on the capabilities of the ion source in use and on the maximum photon energy available (e.g., 340~eV at the ALS \cite{Covington2002}, 1000~eV at SOLEIL \cite{Gharaibeh2011a}, and 3000~eV at PETRA\,III \cite{Schippers2014}). The most highly charged ion investigated at a photon-ion merged-beams setup so far is Ce$^{9+}$ \cite{Habibi2009}. Atomic ions with ionization energies below 3000~eV which can, thus, be photoionized in the PIPE setup are, e.g.,  H-like Si$^{13+}$, Li-like Ga$^{28+}$, Na-like Sb$^{40+}$, or Ge-like Pb$^{50+}$ \cite{Kramida2014}. A powerful electron-cyclotron-resonance (ECR) ion source \cite{Trassl2003a} would be capable of delivering such ions with sufficiently high ion currents.

It should be noted that there are competing methods for the experimental investigation of photon-ion interactions in addition to the merged-beams technique.  The dual laser plasma (DLP) method \cite{Kennedy2004a} uses two plasmas which are generated by two consecutive laser pulses. The  UV emission from the first plasma backlights the second plasma that contains the atomic ions of interest. Thus, the DLP technique allows for absorption spectroscopy of multiply charged ions. However it cannot easily provide absolute cross sections, since the target volume and density as well as the intensity of the ionizing radiation field are not sufficiently well under control. More recently, photon-ion interactions have been studied using trapped ions \cite{Thissen2008a,Lau2008,Simon2010a,Bari2011,Milosavljevic2012,Rudolph2013} partly in combination with VUV and X-ray free electron lasers \cite{Epp2007,Bernitt2012}. Generally, it is not easy to obtain absolute cross sections from trapping techniques since the trap inventory can be a mixture of different ions and/or the overlap between the trapped ion cloud and the photon beam cannot be determined with sufficient accuracy. In this respect, the photon-ion merged-beams technique provides better defined experimental conditions than the competing methods.

\section{Ions in space: Photoionization of atomic ions}\label{sec:atomic}

\begin{figure}[b]
\centering{\includegraphics[width=0.5\textwidth]{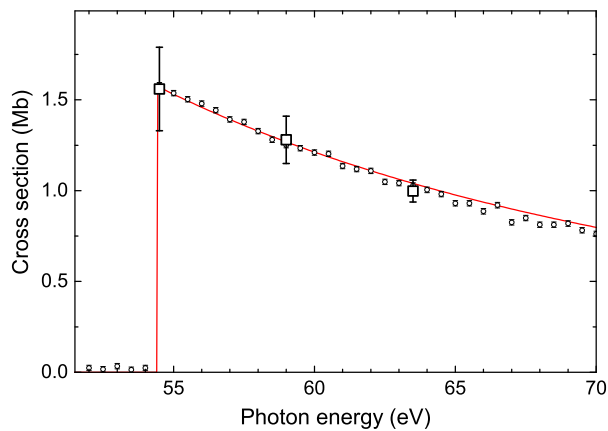}}
\caption{\label{fig:Heabs} Measured (symbols) and calculated  [Equation~(\ref{eq:Stobbe}) with $Z=2$, red full line] cross sections for the photoionization of H-like He$^+$ ions \cite{Aguilar2003c}. The small symbols are results from a photon energy scan measurement. The corresponding error bars denote statistical uncertainties only. The large symbols represent measured absolute cross sections with the error bars comprising statistical and systematic uncertainties. The cross sections are given in units of megabarns with 1~Mb corresponding to $10^{-18}$~cm$^{2}$.}
\end{figure}

Hydrogen-like ions with just one bound electron are the conceptually simplest atomic systems that can be studied by photoionization. The theoretical cross section for this fundamental process was derived already in the early days of quantum mechanics \cite{Stobbe1930}. For the photoionization of a $1s$ electron it is
\begin{equation}\label{eq:Stobbe}
  \sigma(h\nu) =\frac{512\pi^2\alpha}{3(\varepsilon\!+\!1)^4}\frac{e^{-(4\arctan\sqrt{\varepsilon})/\sqrt{\varepsilon}}}{1-e^{-2\pi/\sqrt{\varepsilon}}}\left(\frac{a_0}{Z}\right)^2
\end{equation}
with the photon energy $h\nu$, the fine structure constant $\alpha \approx 1/137$, the scaled energy $\varepsilon = h\nu/(I_\mathrm{H}Z^2)-1 > 0$, the ionization energy of hydrogen  $I_\mathrm{H}\approx 13.6$~eV, the nuclear charge $Z$, and the Bohr radius $a_0 \approx 0.529\times10^{-10}$~m. The cross section is zero below the ionization threshold  ($I_\mathrm{H}Z^2$, i.e., $\varepsilon=0$) it jumps to a finite value at the threshold and monotonically decreases as the photon energy is further increased. Figure \ref{fig:Heabs} shows the calculated cross section for photoionization of He$^+$ ions together with corresponding experimental data. The theoretical cross section agrees with the measured data points within the experimental uncertainties of $\pm 15\%$.

\begin{figure}
\centering{\includegraphics[width=0.5\textwidth]{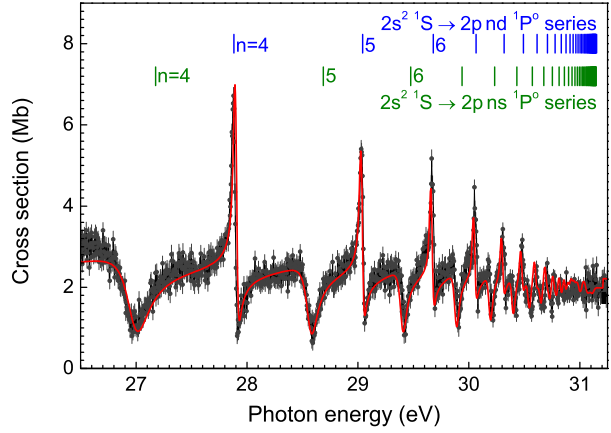}}
\caption{\label{fig:B1rmat} (Modified version of a Figure from Ref.~\cite{Schippers2003b}. \copyright\ 2003 IOP\ Publishing.  Reproduced with permission.  All rights reserved) Photoionization of Be-like B$^+$ ions. The displayed resonances are associated with $2s^2\to 2p\,ns$ and $2s^2\to 2p\,nd$ double excitations from the Be-like $1s^2\,2s^2$ ground configuration. The resonance positions as calculated with Equation~(\ref{eq:Rydberg}) are indicated by the vertical bars. The red full line is a theoretically calculated curve from R-matrix theory.}
\end{figure}

Photoionization becomes more complicated as soon as the target ion has more than one electron. Then, the photon energy can be shared between several electrons leading to multiple excitation or even multiple ionization. In case of (multiple) excitation the photon energy has to match the energy difference between the initial and excited atomic levels. Thus excitation is only possible at discrete photon energies. At these energies, photoionization resonances may be observed if the multiply excited state subsequently decays by autoionization. It should be noted that the excitation of a single outer-shell electron does not lead to the formation of an autoionizing state. Therefore, He$^+$($1s\to n\ell$) single-excitation resonances are not observed in the He$^{2+}$ photo-ion spectrum displayed in Figure~\ref{fig:Heabs}, for example. Figure~\ref{fig:B1rmat} displays the photoionization cross section of Be-like B$^+$($1s^2\,2s^2$), a quasi two-electron system\footnote{The two $1s$ core electrons can safely be considered as mere spectators, since the maximum photon energy of $\sim$ 31.2~eV is well below the B$^+$($1s\to2p$) excitation energy of about 192~eV.}. The observed resonances are due to $2s^2\to2p\,nl$ double excitation with subsequent autoionization into the B$^{2+}$($1s^2\,2s$) ground configuration. Two distinct Rydberg series of resonances can be discerned, the  $2p\,ns\;^2P^o$ and the $2p\,nd\;^2P^o$ series, both converging to the same series limit at $E_\infty \approx 31.15$~eV. To a very good approximation the positions of the $2p\,nl$ resonances can be estimated from the Rydberg formula
\begin{equation}\label{eq:Rydberg}
    E_n = E_\infty-I_\mathrm{H}\frac{(q+1)^2}{(n-\mu)^2}
\end{equation}
where $q=1$ in the case of B$^+$ and the quantum defects  $\mu = 0.408$ and $\mu = -0.087$ for the $2p\,ns\;^2P^o$ and $2p\,nd\;^2P^o$ series, respectively, were obtained from fitting Equation~(\ref{eq:Rydberg}) to the measured resonance positions \cite{Schippers2003b}.

In addition to the resonant ionization channels, there is nonresonant direct ionization of a $2s$ electron leading to a smooth background cross section similar to the one depicted in Figure~\ref{fig:Heabs}. The B$^+$(2s) ionization threshold at 25.15~eV is outside of the photon-energy range of Figure~\ref{fig:B1rmat}. The pathways for $2s$ photoionization of B$^+$ can be written as
\begin{eqnarray}
\textrm{B}^+(1s^2\,2s^2) &\to& \textrm{B}^+(1s^2\,2p\,nl)\nonumber\\
& \searrow & \downarrow\label{eq:pathways}\\
& & \textrm{B}^{2+}(1s^2\,2s)\nonumber
\end{eqnarray}
Both, the pathway via $2p\,nl$ intermediate resonance states and the nonresonant pathway of direct ionization lead from the same initial to the same final state. In such a situation the quantum-mechanical amplitudes associated with both pathways interfere. This interference phenomenon leads to the manifestly asymmetric resonance line shapes in Figure~\ref{fig:B1rmat}. In case of the $2p\,ns$ resonances the interference is largely destructive leading to a reduced cross section in the vicinity of the resonance energies. The (red) full line in Figure~\ref{fig:B1rmat} is the result of a calculation \cite{Schippers2003b} within the nonperturbative theoretical framework of R-matrix theory. The overall agreement with the experimental data is quite satisfactory. The asymmetric resonance line shapes are reproduced very well. The same level of agreement was also found in a comprehensive study on valence-shell photoionization of the heavier Be-like ions C$^{2+}$, N$^{3+}$ and O$^{4+}$ \cite{Mueller2010a}.

As far as absolute cross sections are concerned, the comparison between experiment and theory is often complicated by the fact that not all ions in an ion beam are in their ground level. For example, the $^3P_J$ levels ($J=0,1,2$) in Be-like ions are sufficiently long lived such that they will reach the photon-ion interaction region if they are populated in the ion source. In a photoionization experiment the presence of ions initially in metastable states is revealed by a nonzero ionization signal at photon energies below the threshold for the ionization of the ground level. Usually, the fraction of metastables in an ion beam is unknown. For Be-like  C$^{2+}$, N$^{3+}$, and O$^{4+}$ \cite{Mueller2010a} the metastable fraction amounted to about 50\% as was inferred from the comparison with the corresponding R-matrix results.

In exceptional cases, pure ground-state ion beams can be prepared by passing the ion beam through a gas cell where metastable ions are quenched by collisions \cite{Kjeldsen2002a,Aguilar2003b}. Surface ionization sources produce selected singly charged ions, such as Li$^+$ \cite{Scully2006a}, in their ground level. A more general approach is to store the ions for a sufficiently long time before exposing them to the ionizing photon beam. In a few experimental arrangements, ion traps are used for this purpose \cite{Thissen2008a,Wolf2010}. A draw-back of the trapping technique is that the statistical quality of the results suffers severely from the massive particle losses from the trap during the ion-storage time. The most versatile approach would be the implementation of the photon-ion merged-beams technique at a heavy-ion storage-ring where long ion-storage times can be achieved without excessive beam losses \cite[see, e.g.,][]{Grieser2012}. Its realization would require a major investment and much more floor space than is usually available at synchrotron-radiation facilities. Although heavy-ion storage-rings are presently not available for photoionization studies, the time-inverse process, i.e., photorecombination of ions, is routinely studied at these facilities \cite{Mueller2008a,Schippers2015}. Again, in exceptional cases, the corresponding cross sections can be compared with photoionization cross sections using the principle of detailed balance, and metastable fractions in the photoionization experiment can be inferred from such comparisons on a purely experimental basis \cite{Schippers2002b,Mueller2009a,Mueller2014a}.

Despite the experimental complication concerning metastable ions, conclusive comparisons between experimental and theoretical results (such as the one shown in Figure~\ref{fig:B1rmat}) could be made for a number of atomic ions. Generally, state-of-the-art theoretical methods have been found to be capable of providing cross sections for photoionization of light and few-electron ions with sufficient accuracy for astrophysical purposes. More complex ions are computationally more demanding and approximations must be made in order to keep the calculations tractable. The related uncertainties are difficult to assess and usually only uncovered when theoretical cross sections are compared with experimental data. The large discrepancies between theoretical and experimental data that are revealed in Figure~\ref{fig:FeAstrid} for the photoionization of low-charged iron ions are largely due to the simplifications imposed by the computer technology that was available when the Opacity Project calculations were carried out in the late 1980s.

\begin{figure}
\centering{\includegraphics[width=0.5\textwidth]{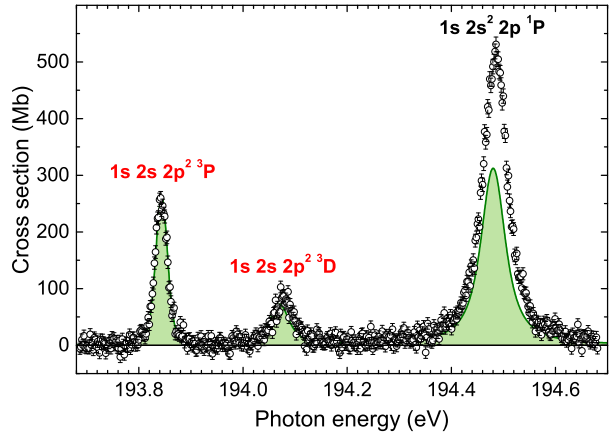}}
\caption{\label{fig:B1Kshell} (Modified version of a Figure from Ref.~\cite{Mueller2014a}. \copyright\ 2014 IOP\ Publishing.  Reproduced with permission.  All rights reserved) Measured (symbols) and calculated (full shaded curve) cross sections for photoionization of singly charged boron ions. The two $1s\,2s\,2p^2$ resonances at 193.84 and 194.07~eV (red labels) are associated with $1s\to 2p$ excitations of $1s^2\,2s\,2p\;^3P$ metastable ions, and the $1s\,2s^2\,2p$ resonance at 194.48~eV (black label) is associated with  $1s\to 2p$ excitation of $1s^2\,2s^2\;^1S$ ground-term ions. The metastable fraction was determined by comparing experimental photoionization and photorecombination cross sections via the principle of detailed balance. Thus, experiment and theory are on independent absolute cross-section scales.}
\end{figure}

Meanwhile, computer technology has greatly advanced and more refined large-scale calculations have become feasible \cite[e.g.][]{Fivet2012, Ballance2015} allowing for a comprehensive treatment of, e.g., K-shell photoionization of light ions. These inner-shell processes are particularly important for opacity calculations \cite{Badnell2003b}. As an example Figure \ref{fig:B1Kshell} shows measured and calculated cross sections for K-shell ionization of B$^+$ ions \cite{Mueller2014a}. Experimental and theoretical resonance positions, widths and strengths agree within the experimental uncertainties with the exception of the strength of the resonance at about 194.5~eV.  This kind of agreement of state-of-the-art theory and experiment can be considered as typical for K-shell ionization of light ions including the astrophysically most important C \cite[e.g.][]{Mueller2009a}, N \cite[e.g.][]{Gharaibeh2011a}, and O \cite[e.g.][]{Bizau2015} isonuclear sequences of ions. Next to hydrogen and helium these elements have the largest abundances in the solar system \cite{Asplund2009}.

On the experimental side, measurements of K-shell photoionization are challenging since the cross sections are rather small except for the few strongest resonances such that only these could be investigated hitherto. Figure~\ref{fig:B1Kshell} provides a typical example. Smaller resonances at higher energies yield increasingly lower signal count rates and their study becomes impractical given the limited amount of beamtime at synchrotron radiation facilities. The experimental means to increase the count rate in a given experimental situation are limited. According to Equation~(\ref{eq:sigma}), one can try to increase either the ion current or the photon flux. The former parameter is limited by the performance of the ion source (which strongly depends on the ion species under investigation) and the latter can only be increased at the expense of resolving power within the limits imposed by the layout of the photon source and the photon beamline.

\begin{figure}
\centering{\includegraphics[width=0.9\textwidth]{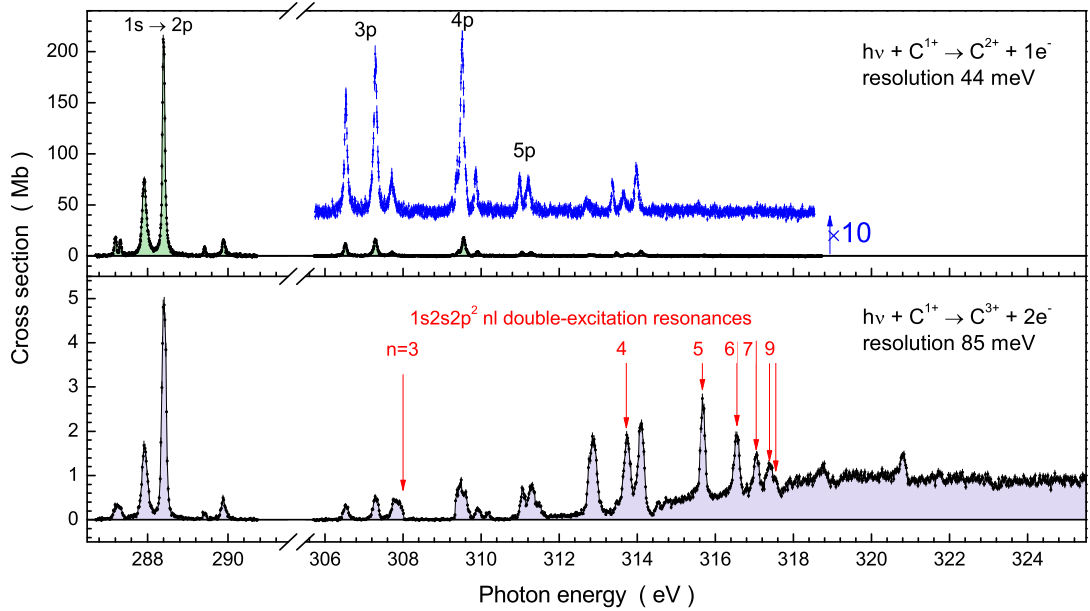}}
\caption{\label{fig:C1K} Measured absolute cross sections for single and double ionization of C$^{+}$ ions \citep{Mueller2015a}. The photon energy axis
has a break in the range where no resonances are expected. The experimental data are shown as small dots with statistical error bars.
They are connected by a solid line with shading. The cross section data for single ionization in the high-energy region were multiplied by
a factor of 10 and displayed again with a vertical offset. Important resonance groups are identified by their configurations.}
\end{figure}

A breakthrough has been achieved recently with the implementation of the photon-ion merged-beams technique at the world's presently brightest 3rd generation synchrotron light source, PETRA\,III in Hamburg, Germany. At the PIPE setup \cite{Schippers2014} (Figure~\ref{fig:PIPE})  record-high photon fluxes are available over an extended photon energy range \cite{Viefhaus2013}. At the same time, the design of the ion beamline was geared towards efficient suppression of background due to stray particles and photons and, thus, towards high signal/background ratios. By these means, the quality of the data could be increased dramatically as is apparent from the comparison of one of the first results from the PIPE setup for C$^+$ \cite{Mueller2015a} (Figure~\ref{fig:C1K}) with a typical result from previous K-shell ionization studies of B$^+$ (Figure~\ref{fig:B1Kshell}).

\begin{figure}
\centering{\includegraphics[width=0.5\textwidth]{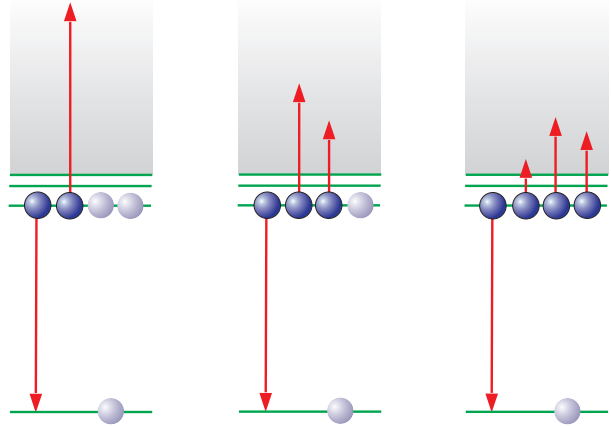}}
\caption{\label{fig:123Auger} Schematic representation of single (left), double (center), and. triple (right) Auger processes in K-shell excited B-like ions (figure taken from \cite{Mueller2015a}).}
\end{figure}

As already mentioned, in these previous studies only the contribution of the strongest resonances to the single ionization channel were investigated. In contrast, the PIPE data for K-shell ionization of C$^+$ ions (Figure~\ref{fig:C1K}) comprise entire Rydberg series of resonances not only in the single but also in the double ionization channel. A prominent additional contribution to the latter channel is nonresonant direct ionization of a $1s$ electron and subsequent autoionization leading to the rise of the cross section beyond the lowest threshold for such a process at about 315~eV. The $1s\,2s^2\,2p\,np$ resonances that are formed via $1s\to np$ excitation of the C$^+$ ion and subsequent autoionization contribute to both ionization channels. For the strongest members of this Rydberg series, i.e., for the $1s\,2s^2\,2p^2\;^2D$ and $1s\,2s^2\,2p^2\;^2P$ resonances at about 288~eV, even triple ionization could be measured. The relative contributions of these resonances to the single, double and triple ionization channels depends on the
respective branching ratios for autoionization of the C$^+$($1s\,2s^2\,2p^2$) resonance states into the C$^{2+}$($1s^2\,2l^2$), C$^{3+}$($1s^2\,2l$), and C$^{4+}$($1s^2$) final states, respectively. In this particular case, the higher charge states can only be formed when the respective Auger processes lead to the simultaneous emission of two or even three electrons (Figure~\ref{fig:123Auger}). The observed ratios of triple- to double- to single-Auger decay rates are of the order of $10^{-4} : 10^{-2} : 1$. In fact, by measuring all cross sections on an absolute scale the rate for the triple Auger process --- representing a genuine four-electron interaction --- could be quantified for the first time \cite{Mueller2015a}.

The relatively light ions discussed up to now have in common that their photoionization resonance structure can be rather easily interpreted in terms of Rydberg series [cf.\ Equation~(\ref{eq:Rydberg})]. The resonance structure of heavier ions with several open electron shells can be very complex due to the associated high density of excited electronic levels. When closely spaced neighbouring resonances overlap the photoionization cross section exhibits broad irregularly shaped features (see, e.g., Figure~\ref{fig:FeAstrid}) instead of individually resolved resonance peaks. Theoretical calculations of these cross sections are demanding because large wave function expansions are required for an accurate description of the resonance levels.

\begin{figure}
\centering{\includegraphics[width=0.5\textwidth]{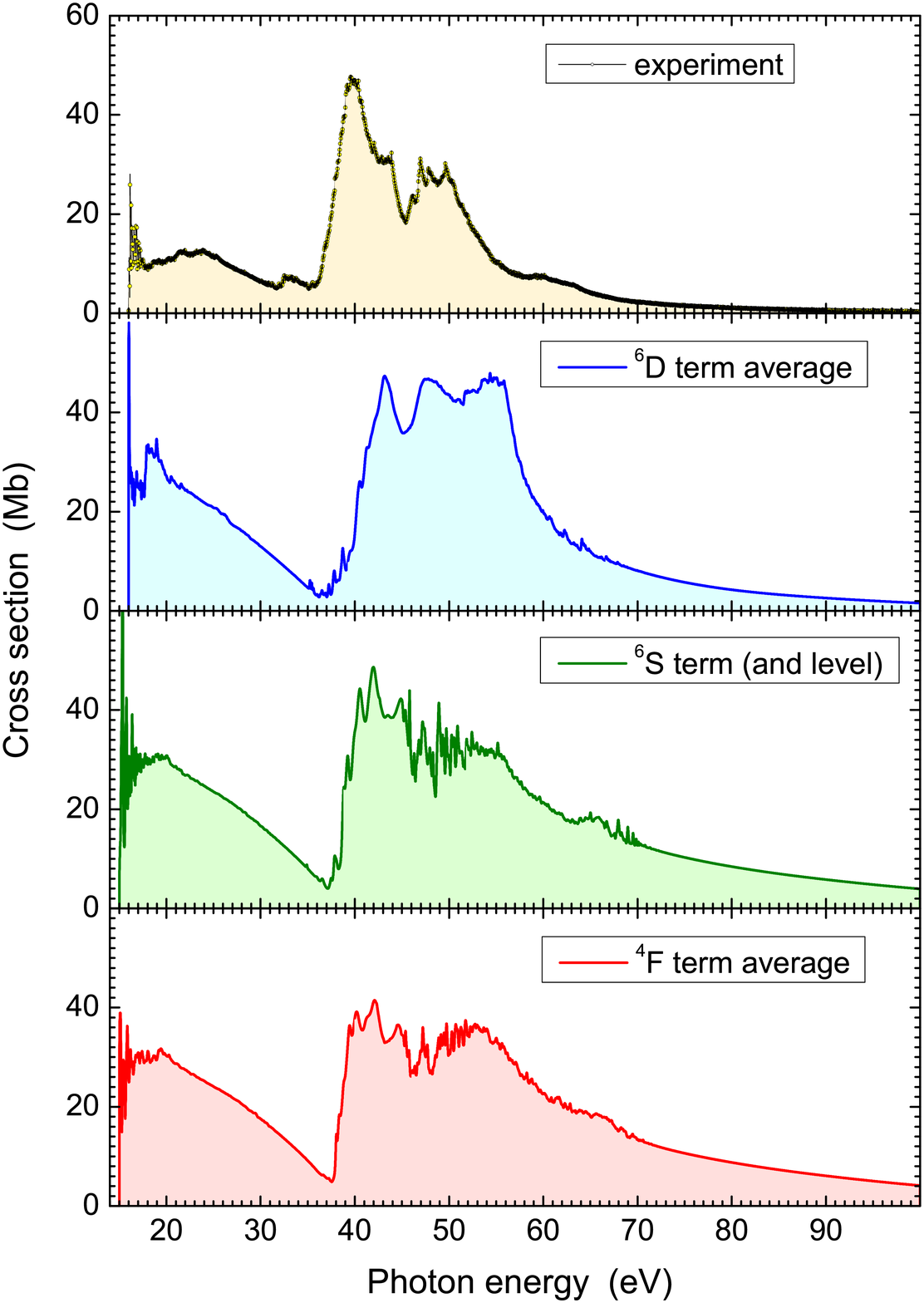}}
\caption{\label{fig:W1PI} (from Ref.~\cite{Mueller2015c}. \copyright\ 2015 IOP\ Publishing.  Reproduced with permission.  All rights reserved) Measured (top panel) and calculated (lower three panels) absolute cross sections for outer-shell photoionization of singly charged tungsten ions.}
\end{figure}

A rather extreme case that has been studied recently by experiment and theory is photoionization of singly charged tungsten ions \cite{Mueller2015c} (Figure~\ref{fig:W1PI}). At present, tungsten receives much attention in atomic physics because of its technological importance in nuclear fusion reactors \cite[][and references therein]{Mueller2015b}. The ground level of W$^+$ with an ionization energy of 16.35~eV  is designated as [Xe]$4f^{14}\,5d^4\,(^5D)\,6s\;^6D_{1/2}$ \cite{Kramida2006}. However, the $5d^4\,6s$ ground configuration is strongly mixed with the neighbouring $5d^3\,6s^2$ and $5d^5$ configurations of the same parity. The associated 13 terms with in total 119 levels are all metastable and were most likely all populated in the primary ion beam. In the photon energy range of up to 245~eV photoionization proceeds via excitation or ionization of a $4f$, $5s$, $5p$, $5d$, or $6s$ electron.

Figure~\ref{fig:W1PI} compares the experimental cross section for photoionization of W$^+$ with the theoretical results for the three lowest terms. Although, the overall shapes of all these cross section curves are very similar it is obvious that a conclusive comparison between theory and experiment cannot easily be made. Over much of the displayed photon energy range, the calculated theoretical cross sections are consistently larger than the experimental cross section by factors 2--3 or more (in particular, at higher energies). At present, the origin of these discrepancies is not clear. It could be conceived that the not yet calculated cross sections for the higher terms might be considerably smaller and that averaging over all terms would bring the theoretical cross section down. Another explanation could be that the wave function expansions are still too limited for an accurate description of the photoionization process. The situation may be similar to electron-ion recombination of complex tungsten ions where statistical theory could successfully be applied for a satisfying theoretical description of the recombination cross section \cite[][and references therein]{Spruck2014a}. This novel theoretical approach has recently been discussed also in the context of photoionization~\cite{Flambaum2015}, but it has not yet been applied to a specific case.

\begin{figure}
\centering{\includegraphics[width=0.5\textwidth]{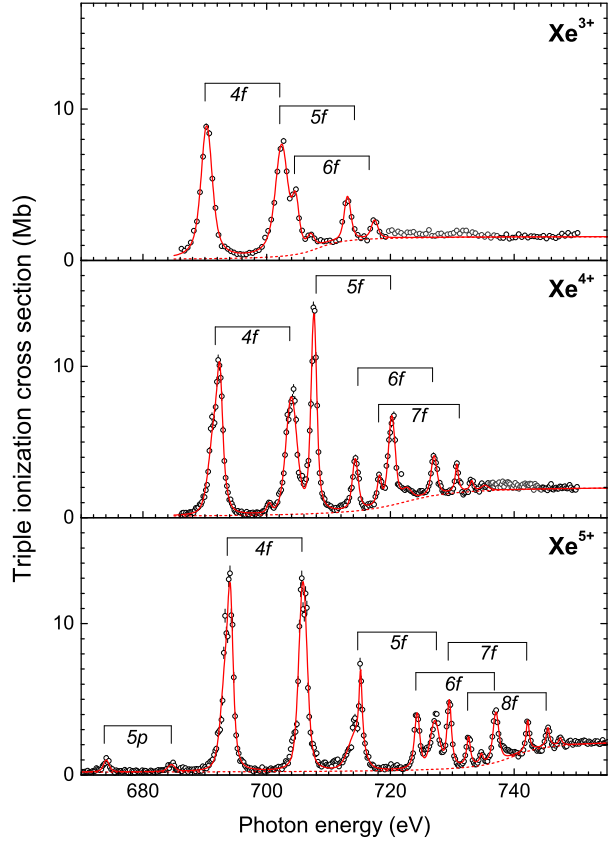}}
\caption{\label{fig:Xe3dhires} (Modified version of a Figure from Ref.~\cite{Schippers2015a}. \copyright\ 2015 IOP\ Publishing.  Reproduced with permission.  All rights reserved) Measured absolute cross sections (symbols) for triple ionization of Xe$^{3+}$, Xe$^{4+}$, and Xe$^{5+}$ ions. The experimental photon energy spread was 160~meV. The full lines are results of Voigt line profile fits to the measured spectra. The dashed lines are the fitted direct ionization (DI) contributions to the measured cross sections. Resonances are labelled by the $n\ell$ subshell to where the $3d$ electron is excited. Because of the fine structure of the $3d$ hole, there are two resonance features for each $nl$, associated with $j=5/2$ (lower resonance energies) and $j=3/2$ (higher resonance energies), respectively.}
\end{figure}

As compared to outer-shell ionization, inner-shell ionization of many-electron ions tends to be less complex since the ionized or excited electron is released from a closed atomic subshell. Accordingly the number of relevant
levels is moderate. For example, Figure~\ref{fig:Xe3dhires} shows experimental cross sections for triple photoionization of low-charged xenon ions in the vicinity of the threshold for the ionization of a $3d$ inner-shell electron \cite{Schippers2015a}. All spectra exhibit a continuous cross section due to direct ionization of the $3d$ electron. However, the strongest features in the spectra are resonances associated with the excitation of the $3d$ electron to an empty $nf$ subshell. There are two corresponding Rydberg series because of the $j$=3/2 -- $j$=5/2 fine-structure splitting (about 13~eV) of the $3d$ core hole.

With increasing charge state increasingly higher $nf$ Rydberg resonances can be excited. This is related to the so called \lq\lq collapse\rq\rq\ of the $nf$ wave functions and can be explained by the peculiar shape of the potential of $f$-electrons. The potential consists of an inner and an outer well separated by a centrifugal barrier. For low nuclear charges, the inner well is too shallow to confine the $nf$ electrons, so that they are mainly localized in the outer well where there is practically no overlap with the $3d$ wave function. When the charge is increased the inner potential well becomes deeper and thus capable of supporting increasingly higher $nf$ wave functions which then have a larger overlap with the $3d$ wave function. Correspondingly, strong resonances associated with $3d\to nf$ excitations are observed for those $nf$ shells which have \lq\lq collapsed\rq\rq\ into the inner well. Although theoretical calculations strongly corroborate this picture \cite{Schippers2015a}, they cannot easily predict the measured triple ionization cross section since the many possible deexcitation pathways (mainly Auger cascades leading to a distribution of final charge states \cite{Schippers2014}) that open up after the initial creation of the core-hole cannot easily be taken into account \cite{Fritzsche2012a}.

\section{Atoms in cages: Photoionization of endohedral fullerene ions}\label{sec:full}

The application of the photon-ion merged-beams technique is by no means limited to atomic ions. In fact, the in-situ mass/charge-selection of ions allows also for experiments with molecular ions \cite[e.g.,][]{Hinojosa2005a} and mass-selected cluster ions. This is an advantage compared to experiments with neutral clusters where the size distribution of clusters often cannot be tightly controlled \cite{Ruehl2003b}. At the same time, sample purity is not an issue since unwanted beam components can be efficiently suppressed by the in-situ mass/charge analysis. The classes of cluster ions that have been most intensely studied so far are charged fullerenes \cite{Scully2005a,Bilodeau2013} and endohedral fullerenes \cite{Mueller2008b,Kilcoyne2010,Phaneuf2013a,Hellhund2015}.

\begin{figure}[b]
\centering{\includegraphics[width=0.4\textwidth]{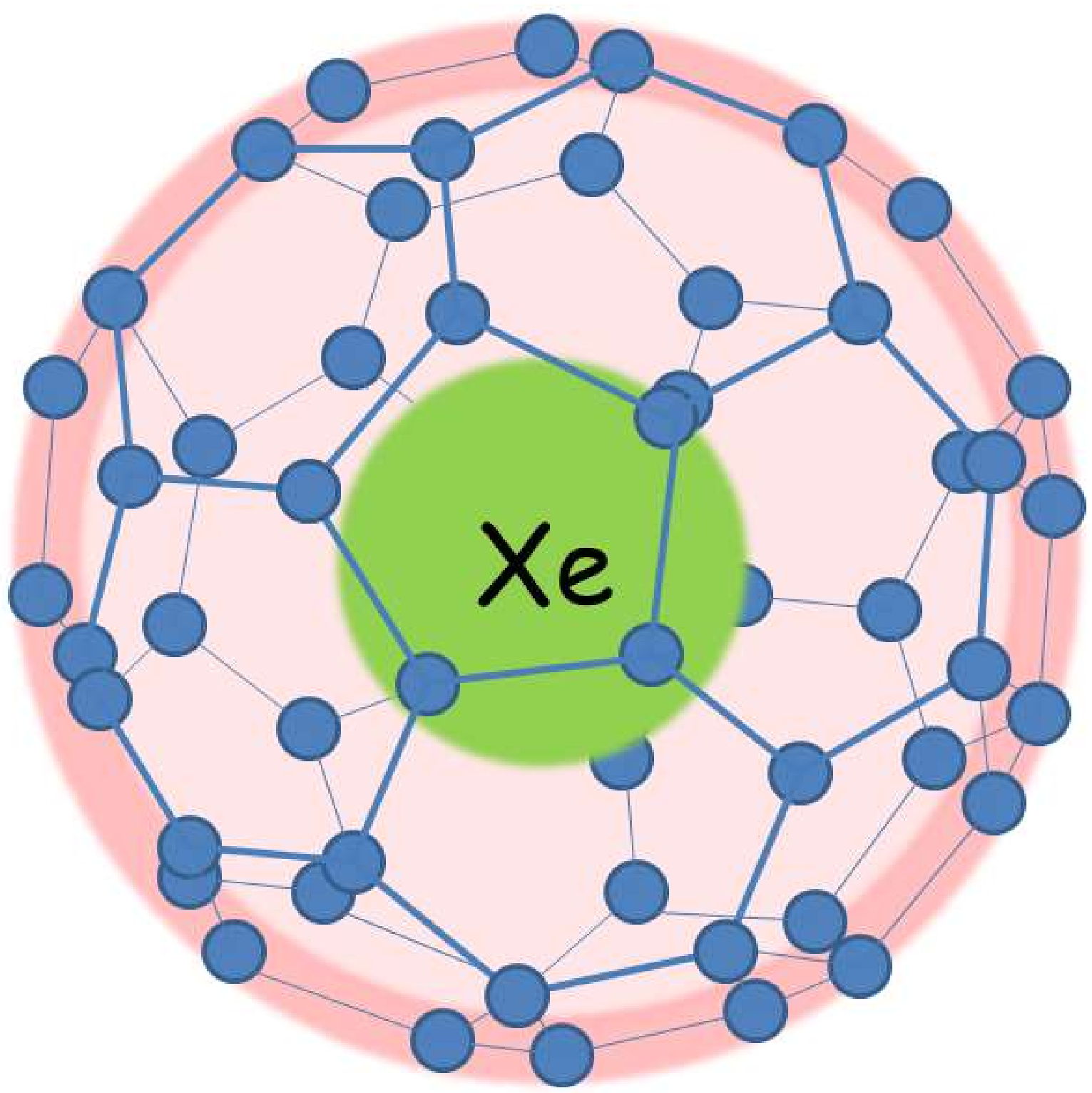}}
\caption{\label{fig:endohedral}Illustration of the endohedral fullerene molecule Xe@C$_{60}$. The Xe atom is centered inside the carbon cage.}
\end{figure}

Endohedral fullerenes are fascinating objects. They consist of a cage of carbon atoms surrounding encapsulated atoms or small molecules (Figure \ref{fig:endohedral}). The fact that the encapsulated atoms or molecules are rather isolated from the surroundings outside of the cage has given rise to interesting ideas for possible applications in many diverse fields such as quantum computing, superconductivity, photovoltaics,  medical imaging, and tumor suppression, to name just a few \cite[][and references therein]{Popov2013}.

\begin{figure}
\centering{\includegraphics[width=0.5\textwidth]{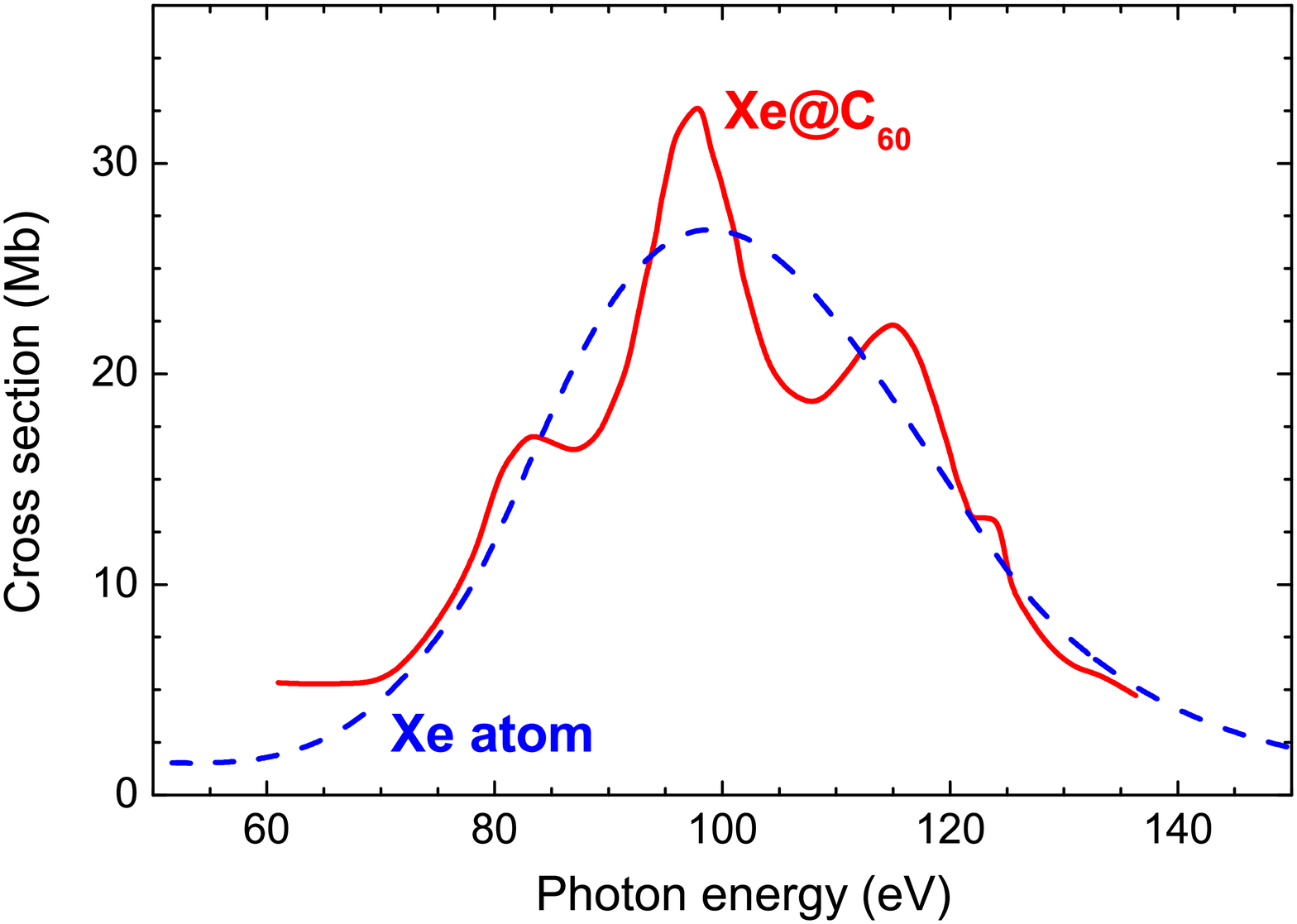}}
\caption{\label{fig:Xe@C60andXe}Photoabsorption cross sections of free Xe atoms \cite[][experiment, dashed line]{West1978}  and of Xe@C$_{60}$ \cite[][theory, full line]{Puska1993} in the energy range of the Xe 4d \lq\lq giant\rq\rq\ resonance.}
\end{figure}

Endohedral fullerenes are also interesting from a fundamental point of view. A particularly intriguing theoretical prediction concerns photoabsorption by encaged atoms. Figure \ref{fig:Xe@C60andXe} shows the experimental photoabsorption cross section of a free Xe([Kr]$\,4d^{10}\,5s^2\,5p^6$) atom (dashed line) in the photon energy range where photoabsorption by the atomic 4d shell yields the dominant contribution to the total cross section. In this energy range the cross section exhibits a \lq\lq giant\rq\rq\ broad resonance. The theoretical prediction \cite{Puska1993} for photoabsorption of a xenon atom encapsulated in a C$_{60}$ fullerene cage is shown as a full red line. Accordingly and somewhat surprisingly, the broad resonance is split into several narrower resonances. These  \lq\lq confinement\rq\rq\ resonances \cite{Connerade2000} can be understood in terms of multi-path interference between electron waves that are emitted directly from the Xe@C$_{60}$ molecule and electron waves that are bouncing on the inner walls of the carbon cage before they are emitted.

\begin{figure}[b]
\centering{\includegraphics[width=0.8\textwidth]{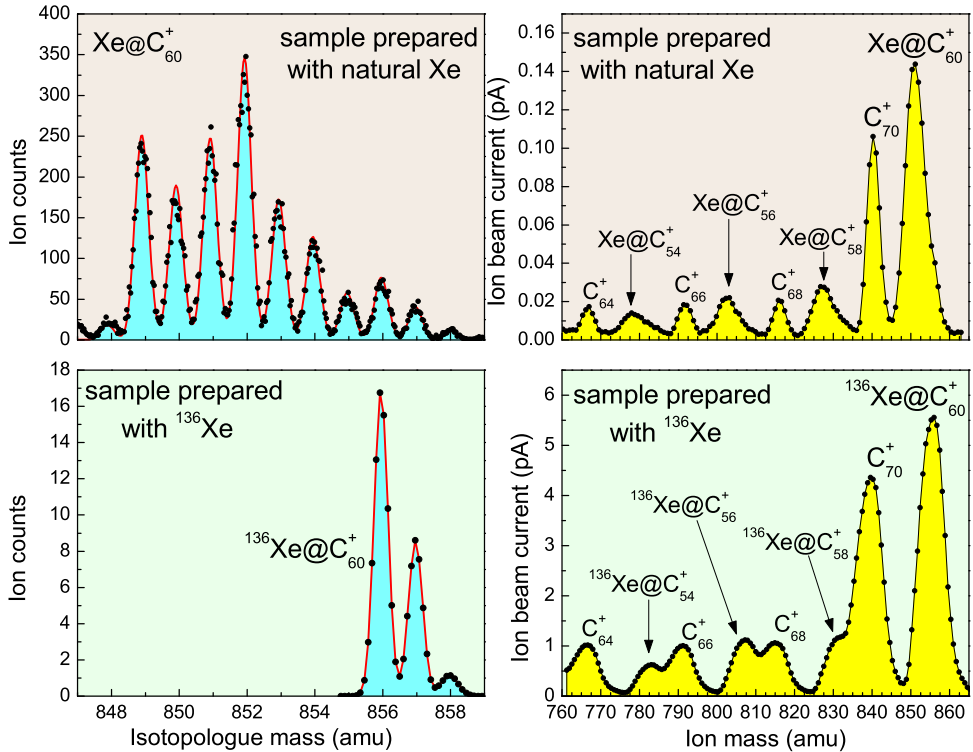}}
\caption{\label{fig:XeC60mass}
Ion beam mass spectra measured at high mass resolution (left panels), and low mass resolution (right panels)~\cite{Phaneuf2013a}. Upper panels are for samples prepared with a natural mixture of Xe isotopes, and lower panels for samples prepared with highly enriched $^{136}$Xe. The samples also contained C$_{70}$ as an impurity. Note that the separation of the Xe@C$_{60}^+$ peak from the C$_{70}^+$ peak could be achieved at a lower mass resolving power in the lower right panel as compared to the upper right panel. The red full line in the upper left panel is a model mass distribution calculated from the natural abundances of C and Xe  isotopes. This distribution is different from the one of natural xenon (Figure~\ref{fig:Xe2mass}) because the distribution of Xe@C$_{60}$ isotopologues is also influenced by the isotope distribution of carbon. The three peaks in the lower left spectrum correspond to $^{136}$Xe@C$_{60}^+$ containing zero, one and two $^{13}$C atoms. }
\end{figure}

Although this phenomenon was investigated theoretically at different levels of sophistication for more than a decade, an experimental verification by a photoabsorption experiment with endohedral fullerenes in the gas phase was achieved only recently \cite{Kilcoyne2010,Phaneuf2013a}. This long delay is due to the challenging experimental difficulties that had to be overcome before conclusive results could be measured. The most direct approach for bringing endohedral fullerenes into the gas phase is evaporation of solid material. Unfortunately, endohedral fullerenes cannot be synthesized in sufficiently large amounts for obtaining vapour targets with a sufficiently high density for detailed photoabsorption studies. Corresponding attempts with rare-earth atoms encapsulated in a C$_{82}$ cage \cite{Mitsuke2005a,Katayanagi2008} have not been very conclusive. Much less material is required for the production of an endohedral fullerene ion beam, in particular, if an ion source is used which can be operated at low vapour pressures.

Figure \ref{fig:XeC60mass} shows mass spectra obtained from samples containing mainly C$_{60}$ as well as traces of larger fullerenes such as C$_{70}$ and of Xe@C$_{60}$. The latter was produced by implantation of Xe$^+$ ions into a layer of 99.95\% pure C$_{60}$ \cite{Kilcoyne2010,Phaneuf2013a}. The samples were evaporated and ionized in an ECR ion source. The displayed mass range comprises Xe@C$_{60}$, C$_{70}$ and their heavier fragmentation products that are created in the ion source, with (endohedral) fullerene fragmentation occurring through the loss of one or several C$_2$ units from the carbon cage. Therefore, only fragments containing even numbers of carbon atoms are observed (right panels of Figure~\ref{fig:XeC60mass}). The strongest peaks correspond to intact C$_{70}^+$ and Xe@C$_{60}^+$ molecular ions. The fact that the yield of Xe@C$_{60}^+$ is comparable to the yield of the impurity ion C$_{70}^+$ shows that the Xe@C$_{60}$ production process was not very efficient. Xenon atoms could be implanted into at best only 1 out of 5000 C$_{60}$ molecules. Nevertheless, an ion beam of pure Xe@C$_{60}^+$ could be produced. In order to be able to optimally separate Xe@C$_{60}^+$ from C$_{70}^+$ isotopically pure $^{136}$Xe was used in the implantation process (lower panels of Figure~\ref{fig:XeC60mass}). The use of xenon with a natural isotope distribution (cf.~Figure~\ref{fig:Xe2mass}) results in a wide mass distribution of several Xe@C$_{60}$ isotopologues as revealed by a corresponding high-resolution mass scan of the Xe@C$_{60}^+$ peak (upper left panel of Figure~\ref{fig:XeC60mass}). The number of isotopologues is greatly reduced in the case of $^{136}$Xe@C$_{60}$  (lower left panel of Figure~\ref{fig:XeC60mass}). These rather involved beam preparation procedures eventually resulted in $^{136}$Xe@C$_{60}^+$ ion currents of up to 5~pA. This corresponded to a number of only $N_\mathrm{ion} = I_\mathrm{ion}L/(ev_\mathrm{ion}) \approx 40$  endohedral molecular ions at a time in the photon-ion interaction region, yielding photo-product count rates of the order of 1~s$^{-1}$ in the cross section maximum  [cf.\ Equation~(\ref{eq:sigma})].

\begin{figure}
\centering{\includegraphics[width=0.5\textwidth]{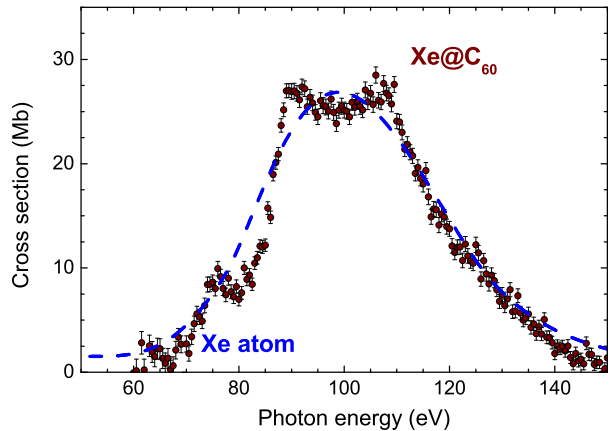}}
\caption{\label{fig:XeC60exp}Experimental results for photoabsorption of Xe@C$_{60}$ \cite[][symbols]{Phaneuf2013a} compared with the experimental results for photoabsorption of free xenon atoms \cite[][dashed line]{West1978}.
}
\end{figure}

Figure~\ref{fig:XeC60exp} shows the cross section that was measured under these conditions in comparison with the cross section for photoabsorption of atomic xenon. The cross section for Xe@C$_{60}$ exhibits clear signatures of confinement resonances albeit less pronounced than theoretically predicted (Figure~\ref{fig:Xe@C60andXe}). There is agreement about the number of resonance maxima (four), but there are differences in resonance positions and sizes. These discrepancies arise due to the rather approximate theoretical treatment of the carbon cage as an effective potential well for the outgoing photoelectron. More refined calculations yield somewhat better but still not perfect agreement between experiment and theory~\cite{Phaneuf2013a}.

In any case, the predicted phenomenon of multi-path interference in the photoionization of an encaged atom has unambiguously been verified by experiment. It should be noted that the visibility of the confinement resonances relies on the central position of the encapsulated atom inside the cage (Figure~\ref{fig:endohedral}). In a less symmetric geometry where the atom is located off-center the interferences are washed out when averaged over the entire solid angle \cite{Korol2011,Chen2014}. For example, no signs of confinement resonances where observed in the photoabsorption of Ce@C$_{82}^+$ \cite{Mueller2008b} where the cerium atom is attached to the inner wall of the carbon cage.

\section{Summary and outlook}

Experimental studies of photoionization of ionized matter requires energetic photons which are readily available from 3rd generation synchrotron light sources. Here we have introduced the photon-ion merged-beams technique that offers a high sensitivity for heavy photo product particles and, thus, partly makes up for the diluteness of ionic targets. So far this technique has been used predominantly for the photoionization of atomic ions, a topic that is strongly driven by atomic data needs in astrophysics and plasma physics. The field has matured in recent years and a satisfying level of theoretical understanding has been reached for few-electron ions. For more complex systems there remain formidable challenges both on the experimental and the theoretical side. The biggest experimental challenge concerns the presence of unknown fractions of metastable ions in the ion beam. Beam storage in a heavy-ion storage ring would provide a solution, but has not been realized so far because it would require a major investment. Recent progress in this direction has been made by the construction of a low-cost electrostatic storage ring for photon-ion interaction experiments, in particular, for the photofragmentation of small molecular ions \cite{Pedersen2015}. However, so far the ion storage times in this device are still rather short. The experiments with endohedral fullerenes have driven the photon-ion merged-beams technique to new limits of sensitivity. Experiments have been conducted successfully with minuscule amounts of sample material. The door to future studies with other types of charged nanoparticles is thus wide open.

\section*{Acknowledgements}

We thank all our collaborators who have participated in the experiments at the Advanced Light Source in Berkeley and at PETRA\,III in Hamburg or who have contributed by providing theoretical calculations. Their names appear in the list of references below. A.M. and S.S. are particulary indebted to Ticia Buhr, Sandor Ricz, and Heinz-J\"urgen Schäfer for their invaluable contributions to the realization of the PIPE setup.  Financial support over the years from the US Department of Energy (DOE, grant nos.\ DE-FG02-03ER15424 and DE-AC03-76SF0098), from the German Ministry of Education and Research (BMBF, grant nos.\ 05KS7RG1 and 05K10RG1), from Deutsche Forschungsgemeinschaft (DFG, grant nos.\ Mu1068/10, Mu1068/20, and Mu1068/22), and through the NATO Collaborative Linkage Grants 950911 and 976362 is gratefully acknowledged.

\newpage

~\\
\parbox{0.25\textwidth}{\includegraphics[width=0.25\textwidth]{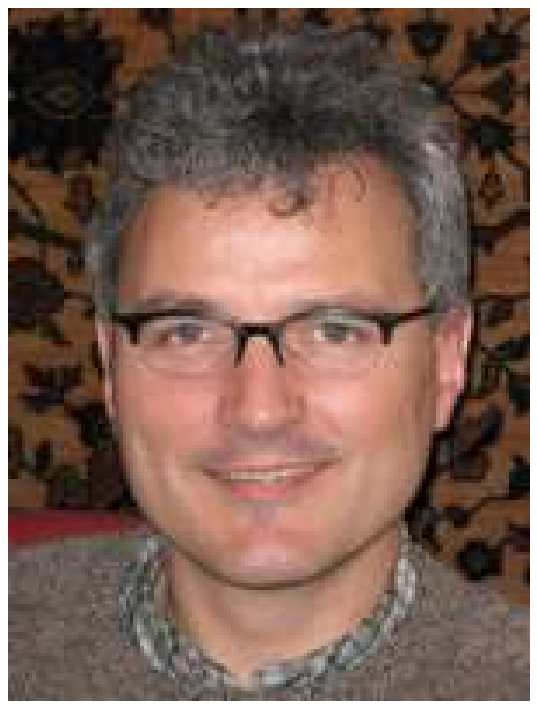}}\parbox{0.05\textwidth}{~}\parbox{0.55\textwidth}{Stefan Schippers received his doctoral degree from the Department of Physics, University of Osnabr\"uck, Germany. Currently he is Adjunct Professor at the Justus-Liebig-University Giessen, Germany. In his research he investigates the interactions of ions with photons, electrons, atoms and solid surfaces, covering topics ranging from fundamental atomic and molecular physics to applications in plasma and astrophysics. Together with Alfred Müller he has been spearheading a scientific collaboration which has built the photon-ion merged-beams experiment PIPE at the synchrotron light source PETRA\,III.}

~\\~\\~\\
\parbox{0.25\textwidth}{\includegraphics[width=0.25\textwidth]{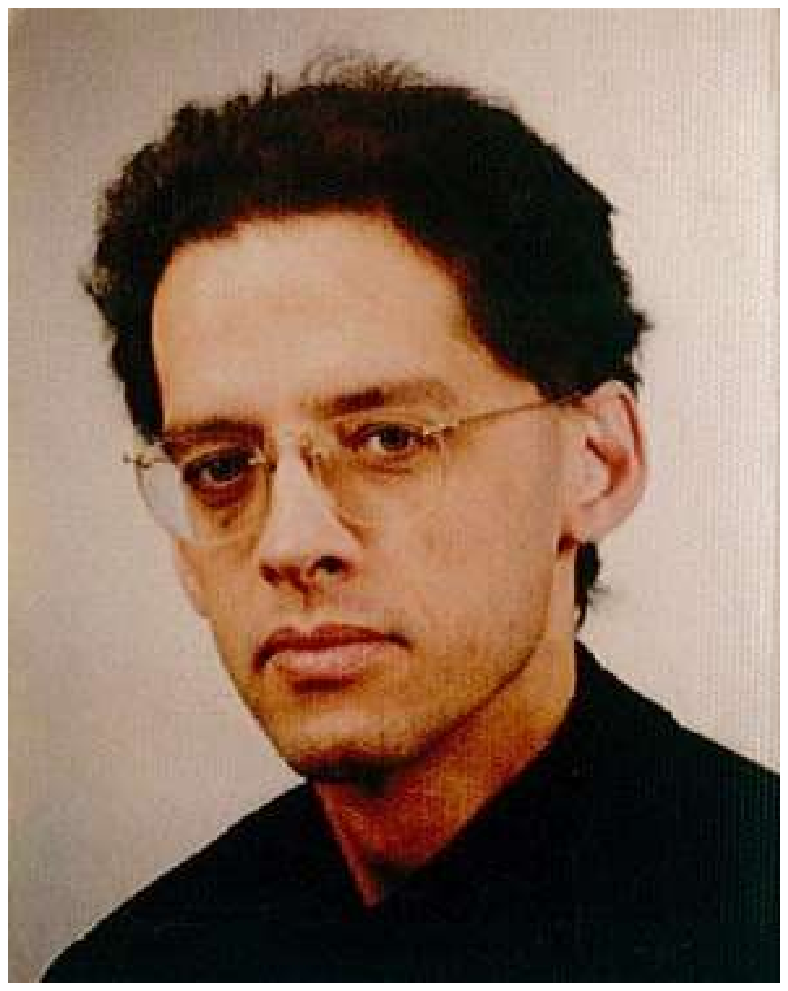}}\parbox{0.05\textwidth}{~}\parbox{0.55\textwidth}{A.~L.~David Kilcoyne received his Ph.D.\ from the Department of Theoretical Chemistry, University of Sydney, Australia. David is a staff member of the experimental systems group at the Advanced Light Source in Berkeley, California, USA. His research interests are in the photoionisation of atomic, molecular and endohedral species and in designing scanning transmission x-ray microscopes for use with soft x-rays from synchrotron radiation sources.}

~\\~\\~\\
\parbox{0.25\textwidth}{\includegraphics[width=0.25\textwidth]{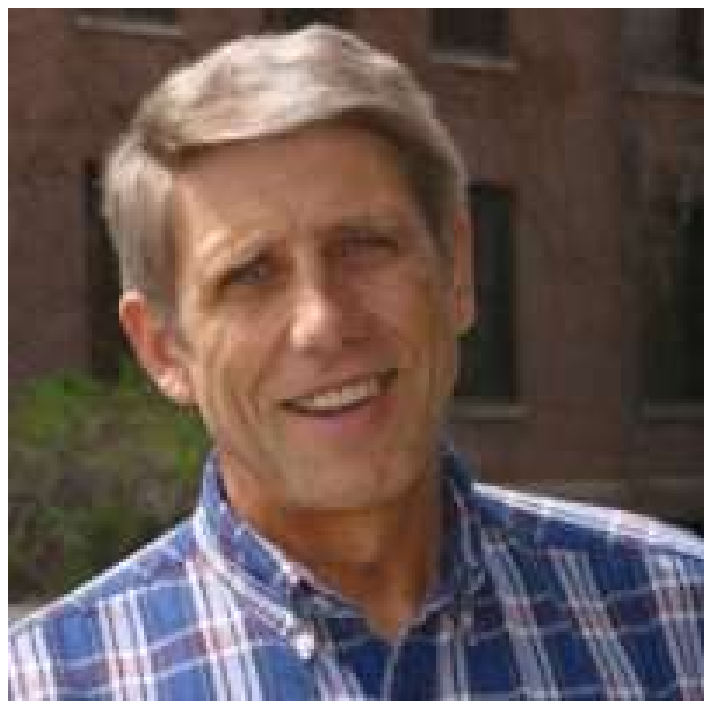}}\parbox{0.05\textwidth}{~}\parbox{0.55\textwidth}{Ronald Phaneuf received a Ph.D.\ in physics from the University of Windsor in Canada and is currently Emeritus Professor of Physics at the University of Nevada, Reno. He had a leading role in the photon-ion research activities at the Advanced Light Source in Berkeley, California. His research interests include the use of merged- and crossed-beams to investigate interactions of atomic and molecular ions with photons, electrons and atoms that are important in high-temperature plasma environments.}

~\\~\\~\\
\parbox{0.25\textwidth}{\includegraphics[width=0.25\textwidth]{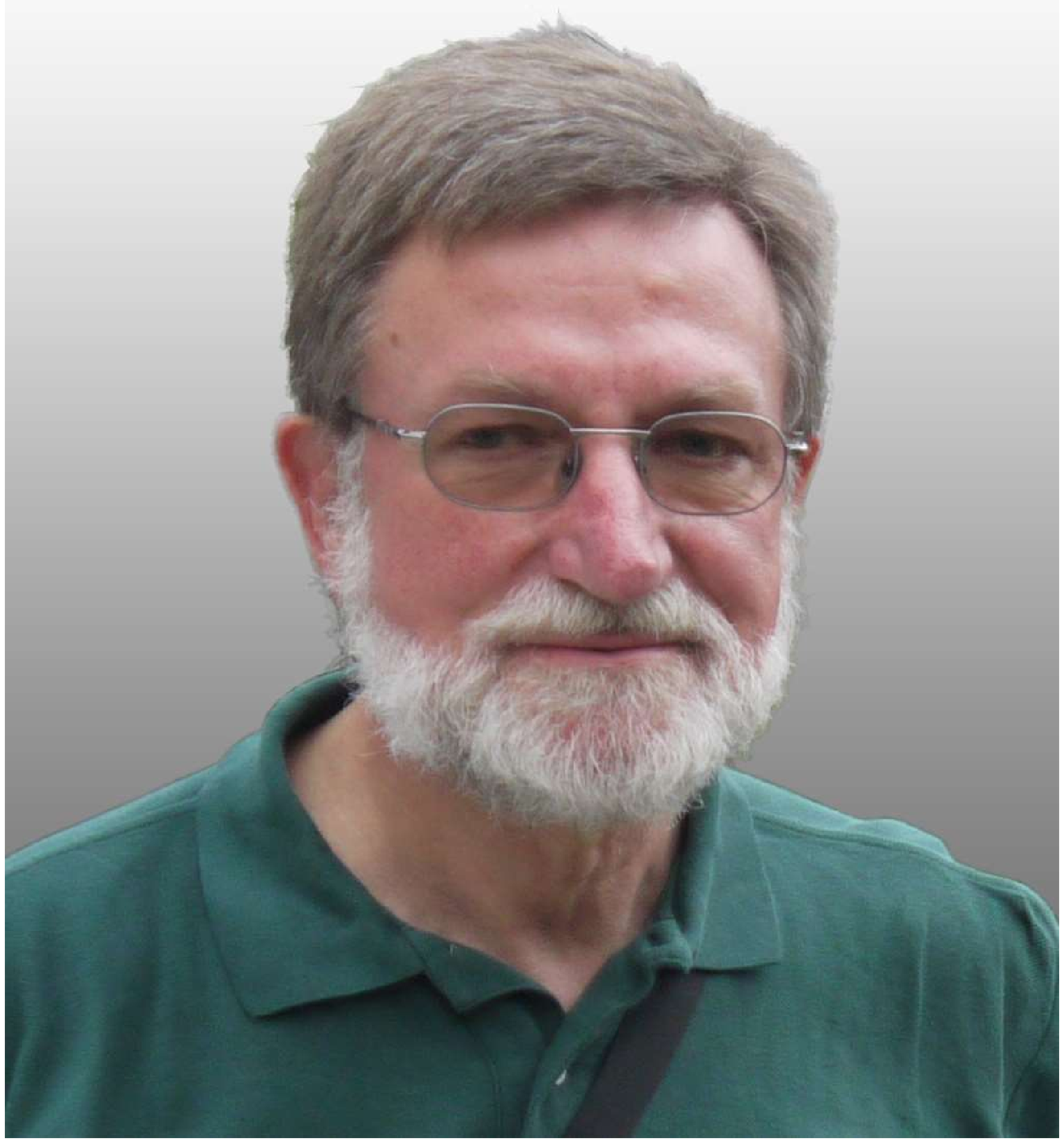}}\parbox{0.05\textwidth}{~}\parbox{0.55\textwidth}{Alfred M\"uller received his doctoral degree from the Department of Physics at Giessen University, Germany. He is a retired Professor of Experimental Physics active in his former research group in Giessen. His scientific interests are in atomic collisions and atomic structure with emphasis on highly charged ions. For most of his research he employed crossed and merged beams techniques. A focus of his work is in high-resolution photoionization, photoexcitation and electron-ion collisional spectroscopy of electrically charged atoms, molecules and clusters.}

\end{document}